\documentclass{sf2a-conf2012}
\usepackage{graphicx}
\usepackage{hyperref}
\usepackage[]{natbib}  
\usepackage[cyr]{aeguill}
\usepackage{epstopdf}

\def\BibTeX{{\rm B\kern-.05em{\sc i\kern-.025em b}\kern-.08em
    T\kern-.1667em\lower.7ex\hbox{E}\kern-.125emX}}
\bibpunct{(}{)}{;}{a}{}{,}  %%%%%%%%%%%%%  A&A bibliography style
\begin{document}

\TitreGlobal{SF2A 2012}

%%-----------------------------------------------------------------
%%      the top matter
%%

\title{Blind decomposition of Herschel-HIFI spectral maps of the NGC 7023 nebula}

\runningtitle{BSS decomposition of HIFI observations}

\author{Olivier Bern\'e}\address{Universit\'e de Toulouse; UPS-OMP; IRAP;  Toulouse, France, CNRS; IRAP; 9 Av. colonel Roche, BP 44346, F-31028 Toulouse cedex 4, France,}

\author{Christine Joblin$^1$}

\author{Yannick Deville$^1$}

\author{Paolo Pilleri}\address{Observatorio Astron\'omico Nacional (OAN), Apdo. 112, 28803 Alcal\'a de Henares (Madrid), Spain}

\author{Jerome Pety}\address{IRAM, 300 rue de la Piscine, 38406 Grenoble Cedex, France}

\author{David Teyssier}\address{European Space Astronomy Centre, Urb. Villafranca del Castillo, P.O. Box 50727, Madrid 28080, Spain}

\author{Maryvonne Gerin}\address{LERMA, Observatoire de Paris, 61 Av. de l'Observatoire, 75014 Paris, France}

\author{Asunci\'on Fuente$^2$}

%% Keep this line, even if the page will be settled afterwards.
\setcounter{page}{237}

%%-----------------------------------------------------------------

\maketitle

%%-----------------------------------------------------------------
%%        The abstract
%% 
%%  Warning!  within the abstract:
%%  - do not use macros. 
%%  - do not use commands like: \cite, \citet, \citep ... etc.

\begin{abstract}

Large spatial-spectral surveys are more and more common in astronomy. This calls for the need of new methods to analyze 
such mega- to giga-pixel data-cubes. In this paper we present a method to decompose such observations into a limited and comprehensive 
set of components. The original data can then be interpreted in terms of linear combinations of these components. The method 
uses non-negative matrix factorization (NMF) to extract latent spectral end-members in the data. The number of needed end-members 
is estimated based on the level of noise in the data. A Monte-Carlo scheme is adopted to estimate the optimal 
end-members, and their standard deviations. Finally, the maps of linear coefficients are reconstructed 
using non-negative least squares. We apply this method to a set of hyperspectral data of the NGC 7023 nebula, obtained recently 
with the HIFI instrument onboard the \emph{Herschel} space observatory, and provide a first interpretation 
of the results in terms of 3-dimensional dynamical structure of the region. 

\end{abstract}

%% Insert the keywords (to appear in the ADS indexing)
%% Keywords must be separated by a comma

%%-----------------------------------------------------------------

Telescopes keep growing in diameter, and detectors are more and more sensitive and made up of an increasing number of pixels. 
Hence,  the number of photons that can be captured by an astronomical instrument, in a given amount of time and at 
a given wavelength, has raised significantly thus allowing astronomy to go {\it hyperspectral}. More and more, 
astronomers do not deal with 2D images, or 1D spectra, but with a combination of both of these media giving
3D data-cubes (2 spatial dimensions, 1 spectral dimension). The PACS, SPIRE  and HIFI instruments, onboard  {\it Herschel} all have a mode allowing spectral 
mapping (e.g. \citealt{vke10, hab10, job10}) in atomic and molecular lines. Owing to its high spectral
resolution, HIFI allows to resolve the profiles of these lines, enabling to study the kinematics of e.g. 
the immediate surrounding of protostars \citep{vds11}, or of star-forming regions \citep{pil12} using radiative
transfer models.

Although such 3D datasets have become common, there is a lack of methods to analyze the
outstanding amount of information they contain.
Classical analysis methods tend to decompose the spectra by fitting them with simple functions
(typically mixture of gaussians) but this has several disadvantages: 1) the assumption made by the use of a given
function usually not based on physical arguments 2) if the number of 
parameters is high, the result of the fit may be degenerate 3) for large datasets and fitting with
nonlinear functions, the fitting may be very time consuming 4) initial guesses must be provided 5)
the spectral fitting is usually performed on a (spatial) pixel by pixel basis, so that the extracted 
components are spatially independent whereas physical components are often present at large
scales on the image.  Alternatively, it is possible to decompose data-cubes using
Principal Component Analysis (e.g. \citealt{ung97, bru02, bru09}). However this, has the disadvantage to decompose data
onto an orthogonal basis, which produces components which are difficult to interpret
directly in physical terms (e.g. spectra with negative values).
An alternative analysis was proposed by \cite{juv96}, which consists in decomposing spectral 
cubes, in this case spectral maps of the pure rotational lines of carbon monoxide (CO), 
into the product of a small number of  spectral components or  ``end members'' and spatial ``abundance'' maps,
with an enforcement of positivity for all the points in the decomposition.
There is no assumption on spectral properties of the components, and hence this can provide deeper insights 
into the physical structure represented in the data, as demonstrated in this pioneering paper. 
This method uses the positivity constraint of the maps and spectra (all their points
must be positive) combined with the minimization of a statistical criterion to derive the 
maps and spectral components. This method is referred to as positive matrix factorization
(PMF, \citealt{paa94}). Although it contained the original idea of using positivity as a constraint
to estimate a matrix product, this work used a classical optimization algorithm. 
Several years later, \cite{lee99} introduced a novel algorithm to perform PMF using 
simple multiplicative iterative rules, making the PMF algorithm extremely fast. 
This algorithm is usually referred to as Lee and Seung's Non Negative Matrix Factorization (NMF
hereafter) and has been widely used in a vast number of applications outside astronomy. Although some
theoretical aspects of this method are still questioned \citep{don04}, this algorithm has proven its efficiency
including in astrophysical applications \citep{ber07}. However, NMF has several disadvantages: 1)
the number of spectra to be extracted must be given by the user and is usually hard to guess 2) the
error-bars related to the procedure are not given automatically.  

In this paper, we present an alternative way of performing a kinematical study of spectral maps, 
using a method which combines NMF to a classification algorithm and a Monte-Carlo analysis. 
This approach tends to discard the standard drawbacks of NMF described above, thus providing a quasi-optimal
decomposition. Here we apply this method to Herschel-HIFI spectral maps of the  [C{\small II}]  and $^{13}$CO(8-7) 
lines at 1900 and 881 GHz respectively. We describe the algorithms we use in Sect. 1 and the method itself in Sect. 2,
and then apply it to the real data.

\section{Definitions and algorithms employed by the method}\label{algo}

\subsection{Mathematical description of the data and aims}

In hyperspectral astronomy, the observed data consist of a 3 D $m\times n \times l$ matrix  $C(p_x, p_y, v)$ where 
$(p_x, p_y)$ define the spatial coordinates and $v$ the spectral index.  We assume that all the points in $C(p_x, p_y, v)$ are 
positive. We call {\it spectrum} each vector $x(p_x, p_y, v)$ recorded at a position $(p_x,p_y)$ 
over the $l$ wavelength points.  The goal of the method that we will describe here is to decompose $C$ 
into the product of a few (typically $<10$) spectra and weight maps using the measured noise in 
the data as the only input to the method and to provide an {\it error} estimation at each point in the 
extracted spectra. The main algorithms we use here are Lee \& Seung's NMF and K-means which we
describe hereafter.

\subsection{Lee \& Seung's NMF in our context} 

We define a new positive 2D matrix of observations $X$, the rows of which 
contain the $m \times n $, {\it spectra} of $C$ arranged in any order. We now assume that each {\it spectrum}, $x$,
is a linear combination of a limited number $r$ (with $r\ll m \times n$,) of unknown {\it source
spectra}, i.e.

\begin{equation}
\label{specr}
x(p_x, p_y, v)= \sum_{i=1}^r a^{i}(p_x, p_y) {s^i(v)}+n(p_x,p_y,v) ,
\label{NMF_def}
\end{equation}
where $s^i(v)$ are {\it source} spectra, $i$ is the {\it source} index, $a^i(p_x, p_y)$ are the unknown ``weight" coefficients 
and $n(p_x,p_y, v)$ is additive noise. This can be re-written in the following matrix form:

\begin{equation}
X=A S +N,
\label{matrix_def}
\end{equation}
where $A$ is the $ m \times n \times r$ matrix of unknown coefficients of the linear combinations
and $S$ is an $r \times l$ matrix, the rows of which are the {\it source} spectra and
$N$ is the noise matrix.
This is a typical blind source separation (BSS) problem \citep{car98}, and can be solved
using multiple methods (e.g. \citealt{lee99}, \citealt{hyv99}, \citealt{gri06}). Here, we use Non-Negative matrix 
factorization \cite{lee99} that is applicable because $A$ and $S$ are positive.
The objective is to find estimates of $A$ and $S$, respectively $W$ and $H$
so that 
\begin{equation}\label{approxg}
X \approx W H.
\end{equation}

This is done by adapting the non-negative matrices $W$ and $H$
so as to minimise the divergence $\delta(X|WH)$, defined as
\begin{eqnarray}
\label{def2}
\delta(X|WH) & = &  \sum_{ij}(X^{ij}\log \frac{X^{ij}}{(WH)^{ij}}\\
				&     & -X^{ij}+(WH)^{ij}),  
\end{eqnarray}
where the exponents $i$ and $j$ respectively refer to the row and column indexes of the matrices.
The algorithm used to achieve the minimization of the divergence is based on the iterative
update rule

\begin{eqnarray}
%&H_{i,j} \leftarrow H_{i,j} \frac{\sum_{u}W_{ui}X_{uj}/(WH)_{uj}}{\sum_{v}W_{vi}},& \nonumber\\
%&W_{i,j} \leftarrow W_{i,j} \frac{\sum_{u}H_{ju}X^{i}/(WH)^{i}}{\sum_{j}H^{j}} .&
&H_{a\mu} \leftarrow H_{a\mu} \frac{\sum_{i}W_{ia}X_{i\mu}/(WH)_{i\mu}}{\sum_{k}W_{ka}}, & \nonumber\\
&W_{ia} \leftarrow W_{ia} \frac{\sum_{\mu}H_{a\mu}X_{i\mu}/(WH)_{i\mu}}{\sum_{\nu}H_{a\nu}}. &
\end{eqnarray}

Divergence is non increasing under their respective update rules,
so that starting from random $W$ and $H$ matrices, the algorithm will converge towards
a minimum for $\delta$. This provides the matrix $H$ containing the $r$ estimated {\it source}
spectra $h_i$ (in the following we refer to these as simply ``{\it source} spectra''). 
The convergence criterion that we have used here measures the evolution of the divergence
as a function of the iteration step. We have assumed that convergence is reached when

\begin{equation}
1-\frac{\delta^{i+1}}{\delta^{i}}<\kappa,
\end{equation}
where $\kappa$ is small, typically 0.0001.

%is defined according to the power lying in the observational noise by
%\begin{equation}
%\kappa=
%\end{equation}.
%This of course is a theoretical definition and in reality we will only have access $\hat{\kappa}$,
%an estimation of $\kappa$. There are several ways to derive $\hat{\kappa}$ from the observations
%which will be described in a subsequent paper (BernŽ et al. in prep.).

\subsection{K-means}

{\it K-means} is a standard unsupervised classification method wich aims to partition
a set of vectors into $k$ sets $\sigma=\{\sigma_1,\sigma_2, \sigma_3 \ldots \sigma_k\}$ so that within
each set, a distance is minimized. In our case, this distance is 1 minus the correlation
coefficient.
Formally this reads:
\begin{equation}
argmin \sum_{i=1, k} \sum_{h_j \in  \sigma_i} 1-corr(h_j, \mu_i),
\end{equation}
where $corr$ stands for correlation coefficient and $\mu_i$ is the
mean spectrum in cluster $\sigma_i$. Said in a simple way, {\it K-means} as
defined here forms clusters that maximize the correlation between vectors 
within each cluster. The algorithm used to perform {\it K-means} is the one
provided in Matlab.

\section{Architecture of the method}\label{archi}

This section describes how we use NMF and {\it K-means} as well as a Monte-Carlo (MC) analysis to
 build our method. The three distinct steps of this method are:
\begin{itemize}

\item{identification of the number of {\it source} spectra based on the difference between the estimated power of noise and reconstruction residuals}
\item{estimation of the {\it source} spectra using NMF and errors using MC analysis,}
\item{reconstruction of the weight maps using non-negative least squares.}

\end{itemize}

These steps are described in the following sections.

\subsection{Identification of the number of {\it source} spectra}

The identification of the number of sources in our method relies on the estimation 
of the norm of the noise matrix $N$ in the data. In real observations $N$ is unknown. However,
it is often possible to get an accurate estimation of the power of the noise in
the data (this will be described in a subsequent paper, Bern\'e et al. in prep.).
Let us consider $\nu_{rms}$, which is an estimation of the 
Frobenius norm of $N$:
\begin{equation}
\nu_{rms}\sim\|N\|_{F}.
\end{equation}

In order to identify the number of {\it source} spectra, NMF as described above
is applied on $C$ starting with the minimum number of {\it source} spectra $r=2$. 
Once NMF has converged, the matrix of approximated observations $\hat{X}$ is obtained by:
\begin{equation}
\hat{X}=W H.
\end{equation}
The norm of residuals (i.e. the difference between original and reconstructed observations)
is calculated by:
\begin{equation}
\Psi=\|\hat{X}-X\|_F.
\end{equation}
If
\begin{equation}
\Psi-\hat{\nu}_{rms}\leq 0,          \label{ineq}
\end{equation}
then the number of sources is $r=2$. If the above inequality
is not verified then the algorithm retries for $r=r+1$ and
such until Eq.~(\ref{ineq}) is verified. Note that we do not bring the
theoretical proof that $\Psi$ is decreasing when $r$ increases, however,
empirical tests on several data-sets (including the one presented in this
paper in Sect. \ref{appli}) show that this is the case.

\begin{figure}
\centering
\includegraphics[width=15cm]{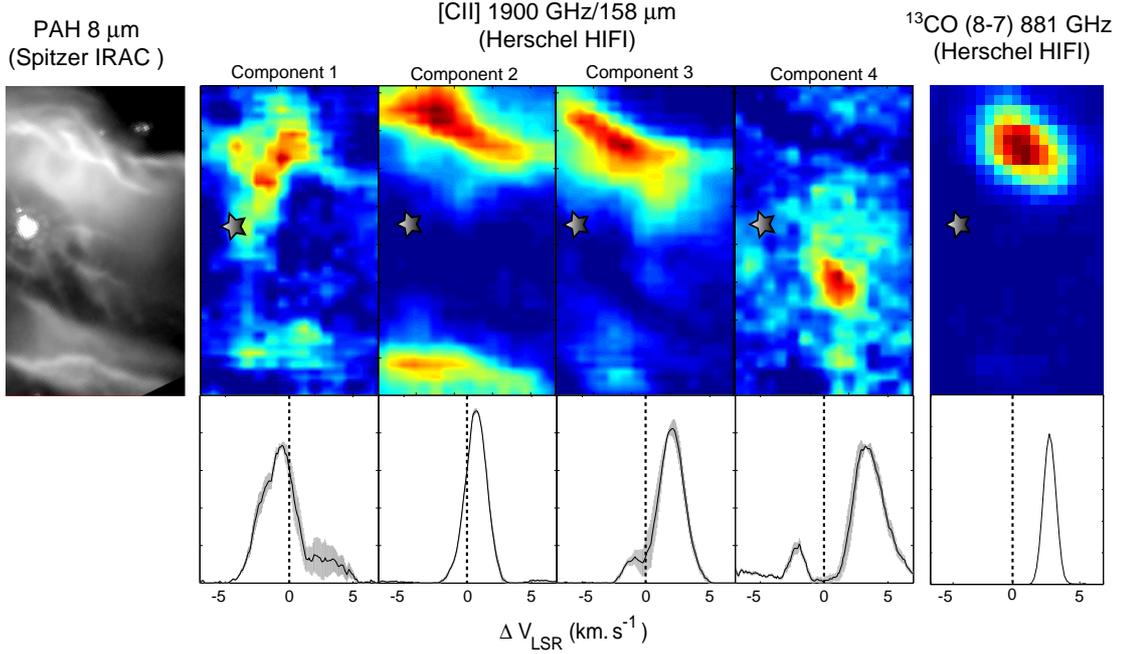}
\caption{ Illustration of the results of our method on the \emph{Herschel}-HIFI data of 
the NGC 7023 nebula. On the left, the \emph{Spitzer}-IRAC images shows the structure of the nebula at
mid-infrared wavelengths (polycyclic aromatic hydrocarbon emission). The right part of the figure shows the weight maps (upper panel) and spectral 
components (lower panel) extracted using the NMF-based method described in this paper.
Color-scale and intensities in the spectra are in arbitrary units (due to the scale uncertainty inherent to blind 
signal separation methods such as NMF). The error intervals ($\epsilon(v)$) as estimated by the method for
 each \emph{source} spectrum is shown with the gray envelope (which is very small in the case of 
 the $^{13}$CO line). The dashed line indicates $\Delta~V_{LSR}=0$.}
\label{fig2}
\end{figure}

\subsection{Monte-Carlo estimation of the {\it source} spectra and empirical standard deviation}

Once the number of sources $r$ has been identified we can run NMF with
this value of $r$ fixed, for $p$ trials, with different initial random matrices for each trial.
$p$ is typically 100. For each value trail a set of $r$ source spectra are identified. The total number
of obtained spectra at the end of this process is hence $p\times r$.
The algorithm uses $K-means$ to form $r$ sets $\{\sigma_1, \sigma_2, \ldots \sigma_r\}$
each set  containing $k_1, k_2, \ldots, k_r$ spectra. Each final $source$ spectrum is obtained by 
averaging the $k_n$ spectra in each set $\sigma_n$.  The rows of a new matrix called $H_*$
contain these $r$ averaged vectors. The standard deviation at each
wavelength $v$ for a given {\it source} spectrum is obtained by 
\begin{equation}
\epsilon_n(v)=\sqrt{\frac{1}{k_n-1}\sum_{i=1}^{k_{n}}(h_i(v)-\bar{h}(v))^2},
\end{equation}
where $h_i(v)$ is the value of the source spectrum in a set at a given wavelength.
$\bar{h}$ denotes the average of $h_i(v)$ in the cluster.  $\epsilon$ is therefore
the empirical sample standard deviation of the value of a point in a $source$ spectrum 
within its cluster.

\subsection{Reconstruction of the weight maps}

The matrix of weights $W_*$ is reconstructed by minimizing
\begin{equation}
\| X-W_*H_*\|.
\end{equation}
This is done by using the classical non-negative least square algorithm.
In the present case we have used the version of this algorithm
provided in Matlab.

\section{Application to Herschel observations}\label{appli}

\subsection{Data}\label{data}

The \emph{Herschel} space telescope is an infrared observatory combining a 3.5 meter telescope and 3 instruments.
Here we have used data from the Heterodyne Instrument for Far Infrared (HIFI) which allows spectral mapping at high
spectral resolution ($v/ \Delta v \sim 10^6$) and high angular resolution ($\sim 10-40''$) between 
480-1250 GHz, and 1410-1910 GHz, as part of the open time
program ``Physics of gas evaporation at PDR edges'' (PI C. Joblin). We have observed
the NGC 7023 nebula \citep{fue96}, where a massive star has blown a cavity inside the cloud where it formed. 
It is illuminated by the young Be star HD200775. With HIFI we have observed the fine structure line emitted by the atomic carbon 
ion (C$^{+}$) at 1900 GHz as well as the pure rotational line of carbon monoxide isotope $^{13}$CO(8-7) at 881 GHz
in spectral mapping. The resulting data consisting in a 3D matrix of 30$\times$52 spatial positions and 201 spectral points
for C$^{+}$ and 18$\times$30 spatial positions and 143  spectral points for the $^{13}$CO(8-7) spectral map.

%. The spectra show that the C$^+$ line profile contains significant
%structure and broadening, which is due to the Doppler effect, induced by the motion of the nebula. For this reason (and as
%done commonly in high resolution astronomical spectroscopy), we do not use standard spectroscopy units e.g.
 %frequency or wavelength but a velocity scale. The zero value on this scale corresponds to the ``Local Standard of Rest'' (LSR),
% basically using the position of the Sun for $v=0$. 

\subsection{Applying the method to our data}

The spectra in this dataset are positive and we assume that non-linear radiative transfer (e.g. re-absorption) effects are not predominant which for the 
C$^+$ and $^{13}$CO(8-7) line in such physical environment is acceptable. Therefore, the proposed method is applicable.
%To estimate $\nu_{rms}$ we extracted as sub cube $C_s$ (small spatial and wavelength range) 
%in $C$ where the spectra contain only noise +continuum. This continuum (or baseline) is subtracted easily using a single slope
%resulting in a cube containing noise only. We use this resulting cube to estimate the power of noise in $C_s$ and scale
 %it to the size of $C$ to obtain $\hat{\nu}_{rms}$. 
 We have applied the method described in Sections \ref{algo} and \ref{archi} to the dataset presented above. 
The whole method, implemented under Matlab runs typically in a few minutes when applied to these data. 
The resulting spectra and weight maps are shown in Fig.~\ref{fig2}. 

\subsection{Preliminary results for C$^{+}$}

Four spectra (and weight maps) are identified 
by the method. Each spatial and spectral component shows distinct features. For clarity,
 the components can be ordered from 1 to 4 based on their kinematic properties in the following way 
 (Fig.~\ref{fig2}): 
\begin{itemize}
\vspace{-0.1cm}
\item{Component 1: lowest peak velocity (about -0.8 km s$^{-1}$ relative to LSR)}
\item{Component 2: mid peak velocity (about +0.8 km s$^{-1}$ relative to LSR)}
\item{Component 3: high peak velocity (about + 2 km s$^{-1}$ relative to LSR)}
\item{Component 4: double peaked, very high (+3 km s$^{-1}$) and very low (-2 km s$^{-1}$) velocity relative to LSR}
\vspace{-0.1cm}
\end{itemize}

\subsection{Preliminary results for $^{13}$CO(8-7)}

For the $^{13}$CO(8-7) line we are only able to find two components, one being a complete noise map,
so in reality it appears as a \emph{unique} $^{13}$CO(8-7) component. This component is shown
in Fig.~\ref{fig2}.

\subsection{Discussion and conclusions}

A preliminary interpretation of these results can be formulated. It seems that we are observing the expansion
of a roughly symmetric cavity, roughly centered at the peak position in the map of component 4. At this position,
we are therefore seeing gas coming towards us and moving away from us at high velocities. The expansion velocity
can be derived from this component by taking half of the velocity difference between the two peaks in the spectrum of component 4,
that is about 2.5 km s$^{-1}$. The other 3 components seem to correspond to the other concentric shells which are 
expanding at the same velocity, but because of the projection effects their resulting absolute velocities as observed with 
\emph{Herschel} are smaller than the expansion velocity of 2.5 km~s$^{-1}$. The radius of the shell can be estimated using
the map of component 2 and taking the distance between the emission peak in the North and the emission peak in the South.
This represents an angular size of 60'', that is $R=3.5~10^{15}$m in physical scale using a measured distance of 430 parsec.   
Using this radius and the expansion velocity we can derive the age of the nebula to be $4.5 \times 10^5$ years.
We also note that the center of the expanding shell does not correspond to the position of the star. This is expected
since this star is known to have a large proper motion, meaning that it must have moved significantly over 
the expansion timescale.The exact origin of this expanding shell needs to be investigated in detail, but it is clear that its expansion is
driven by the massive star that has formed within the molecular cloud \citep{fue96}. Several scenarios are envisaged: winds and outflow from the star,
thermal expansion under irradiation, rocket acceleration of the surrounding cloud due to photo-evaporation etc.
Finally, the fact that the CO component, which traces denser regions, appear as a unique component, is compatible
with the idea that the warmer gas, traced by the C$^{+}$, is subject to much more dynamical structure because of evaporation.
It could also be that there is only one CO component because it traces the edge-on shell which has 
a higher density than the rest of the expanding shell, that can be seen in C$^{+}$. Further interpretation and modeling 
is required to take the best advantage of the results obtained here, however, astronomical observations
in the far infrared at both high angular and high spectral resolutions, can be interpreted in terms of a linear combination 
of few components extracted using an NMF based method.

% Optional acknowledgements
% -------------------------
\begin{acknowledgements}
\end{acknowledgements}

%% The following lines are required when using BibTEX (strongly encouraged!):
\bibliographystyle{aa}  % A&A bibliography style file (aa.bst)
\bibliography{biblio} % your references in file: Yourfile.bib

\end{document}